\documentclass[10pt]{book}
\usepackage{array}
\usepackage{graphicx}
\usepackage{lscape}
\usepackage{epsfig}
\usepackage{amsmath,amssymb}
\usepackage{subfigure}
\usepackage{amsmath,amsthm}

\begin{document}
\renewcommand{\baselinestretch}{1.2}
\newcommand{\ga}{{\alpha}}
\newcommand{\gb}{{\beta}}
\newcommand{\gc}{{\chi}}
\newcommand{\gd}{{\delta}}
\newcommand{\gep}{{\epsilon}}
\newcommand{\gvare}{\varepsilon}
\newcommand{\gga}{{\gamma}}
\newcommand{\gk}{{\kappa}}
\newcommand{\gl}{{\lambda}}
\newcommand{\gm}{{\mu}}
\newcommand{\gn}{{\nu}}
\newcommand{\go}{{\omega}}
\newcommand{\gp}{{\pi}}
\newcommand{\gph}{{\phi}}
\newcommand{\gvp}{{\varphi}}
\newcommand{\gps}{{\psi}}
\newcommand{\gth}{{\theta}}
\newcommand{\gr}{{\rho}}
\newcommand{\gs}{{\sigma}}
\newcommand{\gt}{{\tau}}
\newcommand{\gx}{{\xi}}
\newcommand{\gz}{{\zeta}}
\newcommand{\gE}{{\Upsilon}}
\newcommand{\gGa}{{\Gamma}}
\newcommand{\gL}{{\Lambda}}
\newcommand{\gO}{{\Omega}}
\newcommand{\gP}{{\Pi}}
\newcommand{\gPh}{{\Phi}}
\newcommand{\gPs}{{\Psi}}
\newcommand{\gTh}{{\Theta}}
\newcommand{\gS}{{\Sigma}}
\newcommand{\gX}{{\Xi}}
\newcommand{\gU}{{\Upsilon}}

\newcommand{\fa}{{\bf a}}
\newcommand{\fb}{{\bf b}}
\newcommand{\fc}{{\bf c}}
\newcommand{\fd}{{\bf d}}
\newcommand{\fe}{{\bf e}}
\newcommand{\ff}{{\bf f}}
\newcommand{\fg}{{\bf g}}
\newcommand{\fh}{{\bf h}}
\newcommand{\fj}{{\bf j}}
\newcommand{\fk}{{\bf k}}
\newcommand{\fm}{{\bf m}}
\newcommand{\fn}{{\bf n}}
\newcommand{\fo}{{\bf o}}
\newcommand{\fp}{{\bf p}}
\newcommand{\fq}{{\bf q}}
\newcommand{\fr}{{\bf r}}
\newcommand{\fs}{{\bf s}}
\newcommand{\ft}{{\bf t}}
\newcommand{\fu}{{\bf u}}
\newcommand{\fv}{{\bf v}}
\newcommand{\fw}{{\bf w}}
\newcommand{\fx}{{\bf x}}
\newcommand{\fy}{{\bf y}}
\newcommand{\fz}{{\bf z}}

\newcommand{\fA}{{\bf A}}
\newcommand{\fB}{{\bf B}}
\newcommand{\fC}{{\bf C}}
\newcommand{\fD}{{\bf D}}
\newcommand{\fE}{{\bf E}}
\newcommand{\fF}{{\bf F}}
\newcommand{\fG}{{\bf G}}
\newcommand{\fI}{{\bf I}}
\newcommand{\fJ}{{\bf J}}
\newcommand{\fK}{{\bf K}}
\newcommand{\fL}{{\bf L}}
\newcommand{\fM}{{\bf M}}
\newcommand{\fN}{{\bf N}}
\newcommand{\fO}{{\bf O}}
\newcommand{\fP}{{\bf P}}
\newcommand{\fQ}{{\bf Q}}
\newcommand{\fR}{{\bf R}}
\newcommand{\fS}{{\bf S}}
\newcommand{\fT}{{\bf T}}
\newcommand{\fU}{{\bf U}}
\newcommand{\fV}{{\bf V}}
\newcommand{\fW}{{\bf W}}
\newcommand{\fX}{{\bf X}}
\newcommand{\fY}{{\bf Y}}
\newcommand{\fZ}{{\bf Z}}

\newcommand{\cB}{{\cal B}}
\newcommand{\cE}{{\cal E}}
\newcommand{\cG}{{\cal G}}
\newcommand{\cS}{{\cal S}}
\newcommand{\cF}{{\cal F}}
\newcommand{\cT}{{\cal T}}
\newcommand{\cD}{{\cal D}}
\newcommand{\cI}{{\cal I}}
\newcommand{\cK}{{\cal K}}
\newcommand{\cL}{{\cal L}}
\newcommand{\cM}{{\cal M}}
\newcommand{\cN}{{\cal N}}
\newcommand{\cU}{{\cal U}}
\newcommand{\cW}{{\cal W}}
\newcommand{\cX}{{\cal X}}
\newcommand{\cY}{{\cal Y}}
\newcommand{\cZ}{{\cal Z}}

 \newcommand{\fga}{\mbox{\boldmath $\alpha$}}
  \newcommand{\fgb}{\mbox{\boldmath $\beta$}}
  \newcommand{\fgd}{\mbox{\boldmath $\delta$}}
 \newcommand{\fgg}{\mbox{\boldmath $\gamma$}}
  \newcommand{\fgl}{\mbox{\boldmath $\lambda$}}
  \newcommand{\fgm}{\mbox{\boldmath $\mu$}}
 \newcommand{\fgs}{\mbox{\boldmath $\sigma$}}
 \newcommand{\fgth}{\mbox{\boldmath $\theta$}}
  \newcommand{\fgt}{\mbox{\boldmath $\tau$}}
  \newcommand{\fgz}{\mbox{\boldmath $\zeta$}}
  \newcommand{\fgps}{\mbox{\boldmath $\psi$}}
  \newcommand{\fge}{\mbox{\boldmath $\eta$}}
  \newcommand{\fgga}{\mbox{\boldmath $\gamma$}}
  \newcommand{\fgv}{\mbox{\boldmath $\varphi$}}
  \newcommand{\fgp}{\mbox{\boldmath $\pi$}}
  \newcommand{\fvare}{\mbox{\boldmath $\varepsilon$}}
  \newcommand{\fgD}{{\bf \Delta}}
  \newcommand{\fgG}{{\bf \Gamma}}
  \newcommand{\fgL}{{\bf \Lambda}}
  \newcommand{\fgPh}{{\bf \Phi}}
  \newcommand{\fgPi}{{\bf \Pi}}
  \newcommand{\fgPs}{{\bf \Psi}}
  \newcommand{\fgS}{{\bf \Sigma}}
  \newcommand{\fgTh}{{\bf \Theta}}
  \newcommand{\fgO}{{\bf \Omega}}
  \newcommand{\fgX}{{\bf \Xi}}
  \newcommand{\fgU}{{\bf \Upsilon}}

\newcommand{\Mean}{{\mbox{E}}}
\newcommand{\Cov}{{\mbox{cov}}}
\newcommand{\Var}{{\mbox{var}}}
\newcommand{\Corr}{{\mbox{corr}}}
\newcommand{\diag}{{\mbox{diag}}}
\newcommand{\prob}{{\mbox{Pr}}}

\newcommand{\Nul}{{\bf 0}}
\newcommand{\One}{{\bf 1}}
\newcommand{\Bd}{\B_{\bullet}}
\newcommand{\Xd}{\X_{\bullet}}
\newcommand{\Zd}{\Z_{\bullet}}
\newcommand{\Nor}{N}
\newcommand{\f}{\textbf}
\newcommand{\Real}{{\cal R}}
\newcommand{\Natural}{{\cal N}}
\newtheorem{thm}{Theorem}[section]
\newtheorem{cor}[thm]{Corollary}
\newtheorem{lem}[thm]{Lemma}
\newtheorem{prop}[thm]{Proposition}
\newtheorem{ax}{Axiom}
\theoremstyle{definition}
\newtheorem{defn}{Definition}[section]
\theoremstyle{remark}
\newtheorem{rem}{Remark}[section]
\newtheorem*{notation}{Notation}
\numberwithin{equation}{section}
\newcommand{\thmref}[1]{Theorem~\ref{#1}}
\newcommand{\secref}[1]{\S\ref{#1}}
\newcommand{\lemref}[1]{Lemma~\ref{#1}}
\newcommand{\bysame}{\mbox{\rule{3em}{.4pt}}\,}
\newcommand{\A}{\mathcal{A}}
\newcommand{\B}{\mathcal{B}}
\newcommand{\st}{\sigma}
\newcommand{\XcY}{{(X,Y)}}
\newcommand{\SX}{{S_X}}
\newcommand{\SY}{{S_Y}}
\newcommand{\SXY}{{S_{X,Y}}}
\newcommand{\SXgYy}{{S_{X|Y}(y)}}
\newcommand{\Cw}[1]{{\hat C_#1(X|Y)}}
\newcommand{\G}{{G(X|Y)}}
\newcommand{\PY}{{P_{\mathcal{Y}}}}
\newcommand{\X}{\mathcal{X}}
\newcommand{\wt}{\widetilde}
\newcommand{\wh}{\widehat}
\markright{
} \markboth{\hfill{\footnotesize\rm XIN GAO, DANIEL Q. PU, YUEHUA
WU AND HONG XU }\hfill} {\hfill {\footnotesize\rm TUNING PARAMETER
SELECTION FOR GAUSSIAN GRAPHICAL MODEL} \hfill}
\renewcommand{\thefootnote}{}
$\ $\par \fontsize{10.95}{14pt plus.8pt minus .6pt}\selectfont
\vspace{0.8pc} \centerline{\large\bf TUNING PARAMETER SELECTION
FOR PENALIZED LIKELIHOOD  } \vspace{2pt} \centerline{\large\bf
ESTIMATION OF INVERSE COVARIANCE MATRIX} \vspace{.4cm}
\centerline{Xin Gao, Daniel Q. Pu, Yuehua Wu and Hong Xu}
\vspace{.4cm} \centerline{\it Department of Mathematics and
Statistics, York University, Toronto, Canada} \vspace{.55cm}
\fontsize{9}{11.5pt plus.8pt minus .6pt}\selectfont

\begin{quotation}
\noindent {\it Abstract:} In a Gaussian graphical model, the
conditional independence between two variables are characterized
by the corresponding zero entries in the inverse covariance
matrix. Maximum likelihood method using the smoothly clipped
absolute deviation (SCAD) penalty (Fan and Li, 2001) and the
adaptive LASSO penalty (Zou, 2006) have been proposed in
literature. In this article, we establish the result that using
Bayesian information criterion (BIC) to select the tuning
parameter in penalized likelihood estimation with both types of
penalties can lead to consistent graphical model selection. We
compare the empirical performance of BIC with cross validation
method and demonstrate the advantageous performance of BIC
criterion for tuning parameter selection through simulation
studies.
\par

\vspace{9pt} \noindent {\it Key words and phrases:} BIC;
Consistency; Cross validation; Gaussian graphical model; Model
selection; Oracle property; Penalized likelihood

\par
\end{quotation}\par

\fontsize{10.95}{14pt plus.8pt minus .6pt}\selectfont
\setcounter{chapter}{1}
\setcounter{equation}{0} 
\noindent {\bf 1. Introduction}

A multivariate Gaussian graphical model is also known as a
covariance selection model. The conditional independence
relationships between the random variables are equivalent to the
specified zeros among the inverse covariance matrix. More exactly,
let $X = (X^{(1)},. . . ,X^{(p)})$ be a $p$-dimensional random
vector following a multivariate normal distribution
$N_p(\mu,\Sigma)$ with $\mu$ denoting the unknown mean and
$\Sigma$ denoting the nonsingular covariance matrix. Denote the
inverse covariance matrix as $\Sigma^{-1}=C=(C_{ij})_{1\leq
i,j\leq p}.$ The zero entries $C_{ij}$ in the inverse covariance
matrix indicate the conditional independence between the two
random variables $X^{(i)}$ and $X^{(j)}$ given all other variables
(Dempster, 1972, Whittaker, 1990, Lauritzen, 1996). The Gaussian
random vector $X$ can be represented by an undirected graph
$G=(V,E),$ where $V$ contains $p$ vertices corresponding to the
$p$ coordinates and the edges $E=(e_{ij})_{1\leq i < j \leq p}$
represent the conditional dependency relationships between
variables $X^{(i)}$ and $X^{(j)}.$ It is of interest to identify
the correct set of edges, and estimate the parameters in the
inverse covariance matrix simultaneously.

To address this problem, many methods have been developed up to
date. In general, there would be no zero entries in the maximum
likelihood estimate, which results in a full graphical structure.
Dempster (1972) and Edwards (2000) proposed to use the penalized
likelihood method with the $L_0$-type penalty
$p_{\lambda}(|c_{ij}|)_{i\neq j}=\lambda I(|c_{ij}|\neq 0)$, where
$I(.)$ is the indicator function. Since the $L_0$ penalty is
discontinuous, the resulting penalized likelihood estimator is
unstable. Another standard approach to perform model selection in
Gaussian graphical model is stepwise forward selection or backward
elimination of the edges. However, this approach ignores the
stochastic errors inherited in the multiple stages of the
procedure (Edwards, 2000) and causes the statistical properties of
the method hard to comprehend. Furthermore, the computational
complexity of this greedy search algorithm increases exponentially
with the number of vertices in the graph. Meinshausen and
B\"{u}hlmann (2006) proposed a computationally attractive method
for covariance selection. The proposed method performs the
neighborhood selection for each node and combines the results to
learn the overall graphical structure. It has been shown that this
method is connected to the quadratic approximation of the
loglikelihood with $L_1$ penalty (Yuan and Lin, 2007).
Nevertheless this method performs the model selection and
parameter estimation separately. Yuan and Lin (2007) proposed
penalized likelihood methods for estimating the concentration
matrix with $L_1$ penalty (LASSO) (Tibshirani, 1996). The method
can be implemented through the maxdet algorithm in convex
optimization. However, due to the inherent computational
complexity, the maxdet algorithm can only handle matrices with
small $p$.

Banerjee, Ghaoui and D'aspremont (2007) have proposed a block-wise
updating algorithm for the estimation of inverse covariance
matrix. For each block-wise update, the problem is a
box-constrained quadratic program, which can be solved by an
interior-point procedure. They further showed that the problem
emerges from each step of block-wise update is equivalent to a
linear regression under $L_1$ penalty. Further in this line,
Friedman, Hastie and Tibshirani (2008) proposed the graphical
LASSO algorithm to estimate the sparse inverse covariance matrix
using the LASSO penalty through coordinate-wise updating scheme.
It is the fastest and most convenient algorithm to tackle this
problem up to date. Fan, Feng and Wu (2009) proposed to estimate
the inverse covariance matrix using adaptive LASSO and Smoothly
Clipped Absolute Deviation (SCAD) penalty to attenuate the bias
problem. They employed local linear approximation method (Zou and
Li, 2008) to approximate the LASSO penalty as weighted $L_1$
penalty and the method is implemented through the graphical LASSO
algorithm. The resulted methods with both SCAD and adaptive LASSO
penalties are computationally convenient algorithms leading to
asymptotically unbiased, sparse estimators which possess oracle
property.

In practice, the performance of the penalized likelihood estimator
depends on the proper choice of the regularization parameter. In
this article, we focus on the tuning parameter selection in
penalized likelihood estimation of the sparse inverse covariance
matrix. Wang, Li and Tsai (2007) proposed to use the Bayesian
information criterion (BIC) to select the tuning parameter  for
penalized likelihood method with SCAD penalty, They showed that
BIC with SCAD penalty is able to identify the true model
consistently in the setting of linear regression and partial
linear model. Yuan and Lin (2007) used BIC to select the tuning
parameter with the $L_1$ penalty in the estimation of inverse
covariance matrix. But the consistency of BIC for Gaussian graphic
model has not been investigated. In this article, we establish the
consistency result of the BIC criterion with both SCAD and
adaptive LASSO. We show that if SCAD or adaptive LASSO penalty is
used, the optimum tuning parameter selected by BIC will yield the
graphical structure identical to the true underlying graphical
model with probability tending to one as $n\to \infty.$ We also
compare the performance of BIC with cross-validation method
through extensive simulation studies. We demonstrate that in small
sample size scenario, including the cases when the number of
parameters greatly exceeds the sample size, BIC exhibits
comparable performance as the computationally more intensive
cross-validation method. However, when sample size increases, BIC
consistently outperforms cross validation.

The rest of the article is organized as follows. In Section 2.1 we
formulate the penalized likelihood function for inverse covariance
matrix. In sections 2.2 and 2.3, we discuss the selection of
tuning parameters through the BIC criterion and prove its
consistency in graphical model selection with SCAD and adaptive
LASSO penalty. In section 3, simulation studies are presented to
demonstrate the empirical performance of the tuning parameter
selection with BIC compared with the cross validation method in
small sample size and large sample size scenarios.
\par

\vskip 0.5cm

\setcounter{chapter}{2} \setcounter{section}{1}
\setcounter{equation}{0} 
\noindent {\bf 2.1 Penalized Likelihood Estimation of Inverse
Covariance Matrix}

Given a random sample $X_1,...,X_n$ following a multivariate
normal distribution $N_p(\mu, \Sigma)$, the loglikelihood for
$\mu$ and $C=\Sigma^{-1}$ can be expressed as
\begin{equation*}
\frac{n}{2}\log|C| - \frac{1}{2}\sum_{i=1}^n(X_i-\mu)'C(X_i-\mu),
\end{equation*}
up to a constant not depending on the parameters. The maximum
likelihood estimator of $(\mu, \Sigma)$ is $(\bar{X}, \bar{A})$,
where
\begin{eqnarray*}
\bar A=\frac{1}{n}\sum_{i=1}^n(X_i-\bar X)(X_i-\bar X)'.
\end{eqnarray*}
Assume that the observations are properly centered, then the
sample mean is zero. As $\hat{\mu}$ does not depend on $C,$ we
have $\hat{\mu}=0.$ To obtain the maximum likelihood estimator of
the concentration matrix is equivalent to minimize
\begin{align*}
-\frac 2 n \,\ell(C)=-\text{log}|C|+\text{tr}(C\bar A).
\end{align*}
To achieve sparse graph structure,  penalized likelihood methods
have been proposed in literature and the resulting estimator
$\hat{C}$ should minimize the following objective function:
\begin{eqnarray}
\label{objectivef} Q(C)=-\text{log}|C|+\text{tr}(C\bar
A)+\sum_{i\neq j} p_{\lambda} (|c_{ij}|),
\end{eqnarray}
with $p_{\lambda}$ being some penalty function. Yuan and Lin
(2007) have proposed to use LASSO penalty, $p_{\lambda}
(|c_{ij}|)=\lambda|c_{ij}|.$ Friedman, Hastie and Tibshirani
(2008) proposed the graphical LASSO algorithm by using a
coordinate descent procedure, which is computationally very fast
and guarantees the positive definiteness of the resulting
estimate. As the LASSO penalty increases linearly with the size of
its argument, it leads to biases for the estimates of nonzero
coefficients. To attenuate such estimation biases, Fan and Li
(2001) proposed SCAD penalty. The penalty function satisfies
$p_{\lambda}(0)=0,$ and its first-order derivative is
\begin{align*}
p_{\lambda}'(\theta)=\lambda\{I(\theta\leq \lambda)+
\frac{(a\lambda-\theta)_{+}}{(a-1)\lambda}I(\theta>
\lambda)\},\,\,\text{for}\,\,\theta>0,
\end{align*}
where $a$ is some constant usually set to $3.7$ (Fan and Li,
2001), and $(t)_{+}=t I(t>0)$ is the hinge loss function.

The SCAD penalty is a quadratic spline function with knots at
$\lambda$ and $a\lambda$. It is singular at the origin which
ensures the sparsity and continuity of the solution. The penalty
function does not penalize as heavily as the $L_1$ penalty
function on large parameters. More important advantage of the SCAD
penalty is that the method not only selects the correct set of
edges, but also produces parameter estimators as efficient as if
we know the true underlying graphic structure. Namely, the estimators have the so called
oracle property.

Zou (2006) proposed the adaptive LASSO penalty, which imposes a
weight for each parameter and can be regarded as a weighted
version of the LASSO penalty. In the current setting, the adaptive
LASSO penalty takes the form of $p_{\lambda} (|c_{ij}|)=\lambda
w_{ij}|c_{ij}|,$ with $w_{ij}=1/|\tilde{c}_{ij}|^{\gamma},$ for
 some consistent estimator
$\tilde{C}=(\tilde{c}_{ij})_{1\leq i,j \leq p}$ and some
$\gamma>0.$ As the empirical performance of the results does not
differ much for different $\gamma,$ we follow the conventional
choice of $\gamma=0.5.$

Both SCAD and adaptive LASSO can be efficiently implemented using
the graphical LASSO algorithm. For SCAD penalty, Fan, Feng and Wu
(2009) proposed to use local linear approximation  (Zou and Li,
2008) to approximate the SCAD by a symmetric linear function. The
proposed iterative re-weighted penalized likelihood method
optimizes the objective function at step $(k+1)$ as follows:
\begin{eqnarray}
\label{objectivef1} Q(C)^{(k+1)}=-\log|C|+\text{tr}(C\bar
A)+\sum_{i\neq j} w_{ij}|c_{ij}|,
\end{eqnarray}
with $w_{ij}=p'_{\lambda}(|\hat{c}_{ij}^{(k)}|),$ and
$\hat{c}_{ij}^{(k)}$ denoting the estimates obtained at previous
step. The computation can be implemented by reiteratively using
the graphical LASSO algorithm.

\par
\vskip 0.5cm

\setcounter{section}{2}
\setcounter{equation}{0} 

\noindent {\bf 2.2. Consistency of BIC with SCAD}

In literature, two approaches have been used for the selection of
tuning parameters under the penalized likelihood framework,
including the BIC criterion (Yuan and Lin, 2007) and cross
validation (Friedman, Hastie and Tibshirani, 2008; Fan, Feng and
Wu, 2009). The theoretical investigation of this paper will be
focused on the consistency result regarding the model selection
using BIC criterion under the penalized likelihood framework with
SCAD or adaptive LASSO penalty.

For the tuning parameter $\lambda,$ it is desirable to have a
data-driven method to make the selection automatically. Define the
full graphical model $G_F$ with the full edge set
$E_F=(e_{ij})_{1\leq i < j \leq p}.$ Define an arbitrary graphical
model $G$ with the corresponding edge set $E\subseteq E_F.$ Define
a true model $G_T,$ with the edge set
$E_T=(e_{ij})_{(i,j):c_{ij,0}\neq 0,i<j},$ where $c_{ij,0}$
denotes the null value of the parameter. Define an over-fitted
model $G$ if the corresponding edge set $E\supseteq E_T$ and
$E\neq E_T.$ Define an under-fitted model $G$ with the edge set $E
\nsupseteq E_T.$

In practice, as $\lambda$ is unknown, we search for the optimal
$\lambda$ from the bounded interval $\Omega=[0,\lambda_{\max}],$
for some upper limit $\lambda_{\max}.$ We further assume that the
upper limit $\lambda_{\max} \to 0,$ as $n\to \infty.$ This implies
that the search region shrinks to $0$ as $n$ tends to infinity.
Similar assumption can be found in Wang, Li and Tsai (2007). Given
a tuning parameter $\lambda,$ the penalized likelihood approach
yields the estimated parameters $(\hat{c}_{ij,\lambda})_{1\leq i
\leq j \leq p}.$ The resulting model is denoted as $G_{\lambda}$
with the edge set
$E_{\lambda}=(e_{ij})_{(i,j):\hat{c}_{ij,\lambda}\neq 0}.$ We
define $\Omega_{-}=\{\lambda\in \Omega:E_{\lambda}\nsupseteq
E_T\},$ $\Omega_0=\{\lambda\in \Omega:E_{\lambda}= E_T\},$ and
$\Omega_{+}=\{\lambda\in \Omega:E_{\lambda} \supseteq E_T
\,\text{and}\, E_{\lambda}\neq E_T\}.$ The three subsets of
$\Omega_0,$ $\Omega_{-},$ $\Omega_{+}$ lead to the true, under and
over-fitted models, respectively. Given a $\lambda,$ the
associated BIC criterion is defined as:
\begin{align*}
BIC_{\lambda}=-\log|\hat{C}_{\lambda}|+\text{tr}(\hat{C}_{\lambda}\bar
A)+\frac{\log(n)}{n}\sum_{1\leq i< j\leq p}
I(\hat{c}_{ij,\lambda}\neq 0).
\end{align*}
On the other hand, suppose we know the correct model $G_T$
beforehand and perform the maximum likelihood estimation. Under
$G_T,$ the parameters can be partitioned into two sets:
$C^{(1)}=\{c_{ij}:c_{ij}\neq 0\},$ and
$C^{(2)}=\{c_{ij}:c_{ij}=0\}.$ The resulted maximum likelihood
estimator is denoted as $\hat{C}_{G_T}=(\hat{C}_{G_T}^{(1)},0),$
with $C^{(2)}$ known to be $0.$ The associated BIC criterion is
denoted as
\begin{align*}
BIC_{G_T}=-\log|\hat{C}_{G_T}|+\text{tr}(\hat{C}_{G_T}\bar{A})+\frac{\log(n)}{n}\sum_{1\leq
i< j\leq p} I(c_{ij,0}\neq 0).
\end{align*}

In this subsection, we will focus on the discussion on SCAD
penalty. We first construct a working sequence of reference tuning
parameters $\lambda_n=\log(n)/\sqrt{n},$ which satisfies the
requirement that as $\lambda_n \rightarrow 0,$
$\sqrt{n}\lambda_n\rightarrow \infty.$ Under such working sequence
of tuning parameters, according to Theorem 5.2 in Fan, Feng and Wu
(2009), with probability tending to one, the resulted method will
not only identify the correct set of true edges but also yield
root-$n$ consistent estimators for all the nonzero partial
correlation coefficients. This guarantees the following result:

\begin{lem}
\label{truebic} For SCAD penalty,
$Pr(BIC_{\lambda_n}=BIC_{G_T})\rightarrow 1$ as $n\rightarrow
\infty.$
\end{lem}

\begin{proof}
According to Theorem 5.2 in Fan, Feng and Wu (2009), under the
reference sequence of tuning parameters, we have
$\lim_{n\rightarrow \infty}$
$P(\hat{c}_{ij,\lambda_n}=c_{ij,0})=1.$ It follows that
$\lim_{n\rightarrow \infty}P\bigl(\sum_{i< j}
I(\hat{c}_{ij,\lambda_n}\neq 0)=\sum_{i < j}I(c_{ij,0}\neq
0)\bigr)=1.$ Due to the oracle property, the proposed SCAD
penalized likelihood approach estimates the parameter under the
correct sub-model with probability tending to 1, namely,
$\lim_{n\rightarrow
\infty}P(\hat{C}_{\lambda_n}=\hat{C}_{G_T})=1.$ Then the result of
the Lemma follows.

\end{proof}

Next we will consider the under-fitted model, which is essentially
a misspecified model with at least one of the nonzero parameters
being mistakenly set to zero. Given $\lambda\in \Omega_{-},$ let
$C^{(a)}$ denote $(c_{ij})_{(i,j)\in E_{\lambda}},$ and let
$C^{(b)}$ denote $(c_{ij})_{(i,j)\notin E_{\lambda}}.$  The
penalized likelihood
$\hat{C}_{\lambda}=(\hat{C}^{(a)}_{\lambda},\f 0)$ is the local
minimizer of
\begin{align}
\label{obj} Q_{\lambda}(C)=-\frac 2 n\,
\ell(C^{(a)},0)+\sum_{(i,j)\in E_{\lambda}}p_{\lambda}(|c_{ij}|).
\end{align}
Under the misspecified graphical model, the parameter space is
denoted as $\mathcal{C}^*=\{C|c_{ij}=0, \text{for}\,(i,j)\notin
E_{\lambda}\,\text{and}\,c_{ij}\neq 0, \text{for}\,(i,j)\in
E_{\lambda}\},$ which does not include the true value $C_0.$
According to the asymptotic theory for maximum likelihood
estimation under misspecified model (White, 1982), the maximum
likelihood estimates $\tilde{C}$ will converge to $C^*$ almost
surely where $C^*$ is the unique parameter in the under-fitted
model which minimizes the Kullback-Leibler distance to the true
model, namely
\begin{align}
\label{pseudo} C^*=\text{argmin}_{C \in \mathcal{C}^*}E\{\log
f(X;C_0)-\log f(X,C)\}.
\end{align}
For Gaussian graphical model, the $C^*$ is uniquely defined as the
$-\log f(X;C)$ is strictly convex and so is the $-E \{\log
f(X;C)\}.$ We further partition the pseudo null value
$C^*=(C^{*(a)}, C^{*(b)}),$ with $C^{*(a)}=(C^*_{ij})_{(i,j)\in
E_{\lambda}},$ and $C^{*(b)}=(C^*_{ij})_{(i,j)\notin E_{\lambda}}$
$=\f 0.$ It remains to show that the penalized likelihood
estimator $\hat{C}_{\lambda}$ is also a root-$n$ consistent
estimator to such pseudo null value $C^*.$

\begin{lem}
\label{cstar} Given $\lambda\in \Omega_{-},$ let $C^*$ be defined
as in Equation (\ref{pseudo}). Let the objective function
$Q_{\lambda}(C)$ be defined as in equation (\ref{obj}), with the
penalty being the SCAD function. If $\lambda \rightarrow 0,$ as
$n\rightarrow \infty,$ then there exists a local minimizer
$\hat{C}_{\lambda}$ of $Q_{\lambda}(C)$, such that
$\left\|\hat{C}_{\lambda}-C^*\right\|=O_p(n^{-\frac 1 2}).$
\end{lem}

\begin{proof}
\noindent Consider a constant matrix $u$ with its vectorized form
denoted as $\vec{u}.$ Assume  $u\in \mathcal{C}^*$ and
$||\vec{u}||=M.$ Let $\ell(C)=n/2(\text{log}|C|-\text{tr}(C\bar
A))$. For $n$ large enough,  we have
\begin{align}
\label{dif}
\begin{split}
&n\big(Q_{\lambda}(C^*+\frac{u}{\sqrt{n}})-Q_{\lambda}(C^*)\big)\\
\geq &-2\ell(C^*+\frac{u}{\sqrt{n}})+2\ell(C^*)+n\sum_{(i,j):c^{*(a)}_{ij}\neq 0}\{p_{\lambda}(|c^*_{ij}+\frac{u_{ij}}{\sqrt{n}}|)-p_{\lambda}(|c^*_{ij}|)\}\\
\geq &- 2\ell'(C^*)\frac{\vec{u}}{\sqrt{n}}+\frac 1 {n} \vec{u}^T
 \ell''(C^*)\vec{u} (1+o_p(1))+\\
&n\sum_{(i,j):c^{*(a)}_{ij}\neq
0}\{p'_{\lambda}(|c_{ij}^*|)\frac{u_{ij}}{\sqrt{n}}+
p''_{\lambda}(|c_{ij}^*|)\frac{u_{ij}^2}{n}(1+o(1))\},
\end{split}
\end{align}
with $$\ell'(C^*)=\frac{\partial \ell(C^{(a)},\f 0) }{\partial
C^{(a)}}|_{(C^{*(a)},\f 0)},
$$ and
$$\ell''(C^*)=\frac{\partial^2 \ell(C^{(a)},\f 0) }{\partial^2
C^{(a)}}|_{(C^{*(a)},\f 0)}.
$$

It is known that for $n$ large enough, and $c^*_{ij}\neq 0,$
$p'_{\lambda}(|c_{ij}^*|)=0$ and $p''_{\lambda}(|c_{ij}^*|)=0.$
Furthermore, $C^*$ satisfies $E\{\frac{\partial \log
f(X;C^{(a)},\f 0)}{\partial C^{(a)}}\}|_{(C^{*(a)},\f 0)}=0.$ This
entails $\ell'(C^*)=O_p(\sqrt{n}).$ By standard asymptotic theory,
$\ell''(C^*)=O_p(n).$ By choosing $M$ large enough, the sign of
Equation (\ref{dif}) is completely determined by the second term
of its last line. This implies, for any given $\epsilon
>0,$ by choosing a ball centered around
$C^*,$ with radius $M$ sufficiently large, we have
\begin{align}
P\{\inf_{||u||=M} Q_{\lambda}(C^*+\frac{
u}{\sqrt{n}})>Q_{\lambda}(C^*)\}\geq 1-\epsilon.
\end{align}
This guarantees that the local minimizer $\hat{C}_{\lambda}$ is
root-$n$ consistent for $C^*.$
\end{proof}

The result above is helpful to understand the asymptotic property
of $BIC_{\lambda}$ under an under-fitted model. Concerning an
over-fitted model, some zero-valued parameters are included in the
model to be estimated, which can be regarded as nuisance
parameters. Under an over-fitted model, the parameter space
contains the correct null value of the parameters. Thus the
property of the resulting $BIC_{\lambda}$ can be derived under the
standard likelihood theory under correct model assumption.

\begin{lem}
\label{twocase} If $\lambda_{\max}\rightarrow 0,$ and
$\lambda_{\max}> \log (n)/\sqrt{n}$ as $n \rightarrow \infty,$ and
the penalty is SCAD function, then $Pr(\inf_{\lambda\in
\Omega_{-}\cup \Omega_{+}}BIC_{\lambda}>
BIC_{\lambda_n})\rightarrow 1.$
\end{lem}

\begin{proof}
\noindent First we consider $\lambda \in \Omega_{-}.$  According
to Lemma \ref{cstar}, $\hat{C}_{\lambda}$ is root-$n$ consistent to
$C^*.$ Furthermore, $\ell'(C^*)=O_p(\sqrt{n}),$ and
$\ell''(C^*)=O_p(n).$ We have
\begin{align}
\begin{split}
\ell(\hat{C}_{\lambda})=&\ell(C^*)+\frac{\partial{\ell}}{\partial
C}|_{C^*}(\hat{C}_{\lambda}-C^*)+\frac 1 2
(\hat{C}_{\lambda}-C^*)^T \frac{\partial^2{\ell}}{\partial^2
C}|_{C^*}(\hat{C}_{\lambda}-C^*)(1+o_p(1))\\
=&\ell(C^*)+O_p(1).
\end{split}
\end{align}
By similar argument, $\ell(\hat{C}_{G_T}) =\ell(C_0)+O_p(1).$ By
the weak Law of large numbers,
$$
 \frac{1}{n}\ell(C^*)\xrightarrow[] p E(\log f(X;C^*));
$$
$$
 \frac{1}{n}\ell(C_0)\xrightarrow[] p E(\log f(X;C_0)).
$$
Furthermore, $E(\log f(X;C^*))<E(\log f(X;C_0))$ due to the
Kullback-Leibler inequality. Thus,
\begin{align}
\ell(C_0)-\ell(C^*)=n[E(\log f(X;C_0))-E(\log f(X;C^*))]+o_p(n).
\end{align}
This entails
\begin{align}
\begin{split}
n(BIC_{\lambda}-BIC_{G_T})=&2\ell(\hat{C}_{G_T})-2\ell(\hat{C}_{\lambda})+\log
n
(e_{\lambda}-e_T)\\
=&2\ell(C_0)-2\ell(C^*)+\log n (e_{\lambda}-e_T)+O_p(1)>0,
\end{split}
\end{align}
where $e_{\lambda}=\sum_{i< j}I(\hat{c}_{ij,\lambda}\neq 0),$ and
$e_T=\sum_{i< j}I(c_{ij,0}\neq 0).$

Next consider $\lambda \in \Omega_{+}.$ Define the maximum
likelihood estimator under the true model and under the
over-fitted model as $\hat{C}_{G_T},$ and $\tilde{C}_{\lambda}.$
Note that $\tilde{C}_{\lambda}$ is different from
$\hat{C}_{\lambda},$ as the former is the maximum likelihood
estimate under the submodel $E_{\lambda},$ where the latter is the
penalized likelihood estimate under the full model using $\lambda$
as the tuning parameter. According to the standard asymptotic
theory for the loglikelihood ratio statistic, we have
$2(\ell(\tilde{C}_{\lambda})-\ell(\hat{C}_{G_T}))\sim
\chi^2_{e_{\lambda}-e_T}=O_p(1).$ Furthermore, the penalized
likelihood estimators under the over-fitted model are denoted  as
$\hat{C}_{\lambda}.$ From Theorem 5.2 in Fan, Feng and Wu (2009),
$|\hat{C}_{\lambda}-\tilde{C}_{\lambda}|=O_p(n^{-\frac 1 2}).$
This entails
$\ell(\hat{C}_{\lambda})=\ell(\tilde{C}_{\lambda})+O_p(1).$
Combining the results above, we have
\begin{align}
\begin{split}
n(BIC_{\lambda}-BIC_{G_T})=&-2\ell(\hat{C}_{\lambda})+2\ell(\hat{C}_{G_T})+\log
n
(e_{\lambda}-e_{T})\\
=&-2\ell(\tilde{C}_{\lambda})+2\ell(\hat{C}_{G_T})+\log n
(e_{\lambda}-e_T)+O_p(1)\\
=&\log n (e_{\lambda}-e_T)+O_p(1)>0.
\end{split}
\end{align}
This completes the proof.

\end{proof}

This Lemma implies that the $\lambda$s that fail to identify the
true model yield BIC always larger than $\lambda_n.$ Consequently,
the $\lambda$ value which minimizes the BIC criterion will
identify the true model. Combining the two lemmas above, we
establishes the consistency of the BIC criterion used under the
penalized likelihood framework with the SCAD penalty.

\begin{thm}
If $\lambda_{\max}\rightarrow 0,$ and $\lambda_{\max}> \log
(n)/\sqrt{n}$ as $n \rightarrow \infty,$ then
$Pr(G_{\hat{\lambda}_{BIC}}$ $=G_T)\rightarrow 1,$ where
$\hat{\lambda}_{BIC}$ is the tuning parameter that minimizes the
BIC criterion with the SCAD penalty.
\end{thm}

\par
\vskip 0.5cm \setcounter{section}{3}
\setcounter{equation}{0} 

\noindent {\bf 2.3. Consistency of BIC with adaptive LASSO}

In this section, we focus on the establishment of the consistency
result of BIC with adaptive LASSO penalty. Given any
$a_n$-consistent estimate $\tilde{C},$ namely,
$a_n(\tilde{C}-C_{0})=O_p(1),$ the weights of the adaptive LASSO
 are specified by $w_{ij}=1/|\tilde{C}_{ij}|^{\gamma},$ for some $\gamma>0.$ We first construct a sequence of reference tuning
parameters which satisfies the requirement that as $\lambda_n
\rightarrow 0,$ $\sqrt{n}\lambda_n=O_p(1),$ and $n^{\frac 1
2}\lambda_n a_n^{\gamma}\rightarrow \infty.$  Under such working
sequence of tuning parameters, according to Theorem 5.3 in Fan,
Feng and Wu (2009), with probability tending to one, the resulting
method will not only identify the correct set of true edges but
also yield root-$n$ consistent estimators for all the nonzero
partial correlation coefficients. This guarantees the following
result:

\begin{lem}
\label{truebicad} $Pr(BIC_{\lambda_n}=BIC_{G_T})\rightarrow 1$ as
$n\rightarrow \infty.$
\end{lem}

Next we will consider the under-fitted model in a similar manner
as what we have derived for SCAD penalty.

\begin{lem}
\label{cstarad} Given $\lambda\in \Omega_{-},$ the corresponding
misspecified model is denoted as $G_{\lambda}.$ let $C^*$ be
defined as in Equation (\ref{pseudo}). Let the objective function
$Q_{\lambda}(C)$ defined as in equation (\ref{objectivef}) with
adaptive LASSO penalty. If $\lambda_n \rightarrow 0,$
$\sqrt{n}\lambda_n=O_p(1),$ and $n^{\frac 1 2}\lambda_n
a_n^{\gamma}\rightarrow \infty.$ then there exists a local
minimizer $\hat{C}_{\lambda}$ of $Q_{\lambda}(C)$, such that
$\left\|\hat{C}_{\lambda}-C^*\right\|=O_p(n^{-\frac 1 2}).$
\end{lem}

\begin{proof}
\noindent Consider a constant matrix $u$ with its vectorized form
denoted as $\vec{u}.$ Assume  $u\in \mathcal{C}^*$ and
$||\vec{u}||=M.$ Let $\ell(C)=n/2(\text{log}|C|-\text{tr}(C\bar
A))$. For $n$ large enough,  we have
\begin{align}
\label{difad}
\begin{split}
&n(Q_{\lambda}(C^*+\frac{u}{\sqrt{n}})-Q_{\lambda}(C^*))\\
\geq &-2\ell(C^*+\frac{u}{\sqrt{n}})+2\ell(C^*)+n\sum_{(i,j):c^{*(a)}_{ij}\neq 0}\{p_{\lambda}(|c^*_{ij}+\frac{u_{ij}}{\sqrt{n}}|)-p_{\lambda}(|c^*_{ij}|)\}\\
\geq &- 2l'(C^*)\frac{\vec{u}}{\sqrt{n}}+\frac 1 {n} \vec{u}^T
 \ell''(C^*)\vec{u} (1+o_p(1))+n \lambda_n
\sum_{(i,j):c^{*(a)}_{ij}\neq 0}\{|\tilde{c}_{ij}^*|^{-\gamma}
\frac {u_{ij} }{\sqrt{n}}\text{sign}(c_{ij}^*)\}.
\end{split}
\end{align}

Using similar arguments as in Lemma \ref{cstar}, we have
$\ell'(C^*)=O_p(\sqrt{n}),$ and $\ell''(C^*)=O_p(n).$ Furthermore,
$|\tilde{c}_{ij}^*|^{-\gamma}=O_p(1),$ as $\tilde{c}_{ij}$ is a
consistent estimator of $c_{ij}^*\neq 0.$ Because
$\sqrt{n}\lambda_n=O_p(1),$ the third term is also $O_p(1).$ By
choosing $M$ large enough, the sign of Equation (\ref{dif}) is
completely determined by the second term of its last line. This
implies, for any given $\epsilon
>0,$ by choosing a ball centered around
$C^*,$ with radius $M$ sufficiently large,
\begin{align}
P\{\inf_{||u||=M} Q_{\lambda}(C^*+\frac{
u}{\sqrt{n}})>Q_{\lambda}(C^*)\}\geq 1-\epsilon.
\end{align}
This guarantees that the local minimizer $\hat{C}_{\lambda}$ is
root-$n$ consistent for $C^*.$
\end{proof}

In light of the result above, we are able to study the asymptotic
property of $BIC_{\lambda}$ under the under-fitted model and the
over-fitted model. Let the working sequence of $\lambda_n$ be
defined as above. Following the same argument as in Section 3.1,
we have

\begin{lem}
\label{twocasead} If $\lambda_{\max}\rightarrow 0,$ and
$\lambda_{\max}>\lambda_n,$ as $n \rightarrow \infty,$ then
$Pr(\inf_{\lambda\in \Omega_{-}\cup \Omega_{+}}$ $BIC_{\lambda}>
BIC_{\lambda_n})\rightarrow 1.$
\end{lem}

\begin{thm}
If $\lambda_{\max}\rightarrow 0,$ and $\lambda_{\max}>\lambda_n,$
as $n \rightarrow \infty,$ then
$Pr(G_{\hat{\lambda}_{BIC}}=G_T)\rightarrow 1,$ where
$\hat{\lambda}_{BIC}$ is the tuning parameter that minimizes the
BIC criterion with adaptive LASSO penalty.
\end{thm}

\par
\vskip 0.5cm

\setcounter{chapter}{3}
\setcounter{equation}{0} 
\noindent {\bf 3. Simulation Studies}

Next we conduct simulation studies to investigate the performance
of BIC in penalized likelihood estimation of Gaussian graphical
model. The main focus is to use empirical evidence to support the
consistency result of BIC with SCAD penalty or Adaptive LASSO. We
also compare its performance with cross validation, which is
another commonly used tuning parameter selection method. The
$K$-fold cross-validation method partitions all the samples into
$K$ disjoint subsets and denote the indices of subjects in
$k$-fold by $T_k,$ $k=1,\dots,K.$ The $K$-fold cross-validation
score is defined as:
$$
\text{CV}(\lambda)=\sum_{k=1}^K
n_k(-\log|\hat{C}_{\lambda,-k}|+\text{tr}(\hat{C}_{\lambda,-k}A_k)),
$$
where $n_k$ is the size of the subset $T_k,$
$\hat{C}_{\lambda,-k}$ is the estimated concentration matrix based
on the sample $\cup_{j\not=k} T_j$, and $A_k$ is the sample
covariance matrix calculated on subset $T_k.$ The optimum tuning
parameter $\lambda$ is selected to minimize $\text{CV}.$ In our
simulation, $K$ is set to be $5.$

We simulate three different graphical model structures.

\begin{itemize}
\item Model 1. An AR(1) model is considered with $c_{ii} = 1,$ and
$c_{i,i-1} = c_{i-1,i} =0.5$.

\item Model 2. An AR(2) model is considered with $c_{ii} = 1.5,$
$c_{i,i-1} = c_{i-1,i} = 0.5$ and $c_{i,i-2} = c_{i-2,i} = 0.40$.

\item Model 3. A general sparse graphical model is considered. We
employed the data generating scheme of Li and Gui (2006). To be
more specific, we generate $p$ nodes randomly on the unit and
square and obtained their pairwise Euclidean distance. For each
point, it is connected with an edge to the points with the $3$
smallest distances. For each edge, the corresponding entry in the
inverse covariance matrix is generated uniformly over
$[-1,-0.5]\cup[0.5,1].$ In order to ensure the positive
definiteness of the inverse covariance matrix, the magnitude of
the $i$th diagonal entry is set as twice of the sum of the
absolute values of all the off-diagonal entries on the $i$th row.
\end{itemize}

For each model, we use penalized likelihood methods with SCAD,
adaptive LASSO and LASSO penalties. The tuning parameter for all
three penalties are selected through either the BIC criterion or
the cross-validation criterion. To assess the model selection
performance, we evaluate the sensitivity, specificity, and
Matthews correlation coefficient (MCC) which are defined as
follows:
$$
\text{specificity}=\frac{\text{TN}}{TN+FP},
\text{sensitivity}=\frac{\text{TP}}{TP+FN},
$$
$$
\text{MCC}=\frac{\text{TP}\times \text{TN}-\text{FP}\times
\text{FN}}{\sqrt{(\text{TP}+\text{FP})(\text{TP}+\text{FN})(\text{TN}+\text{FP})(\text{TN}+\text{FN})}},
$$
where $\text{TP},$ $\text{TN},$ $\text{FP},$ $\text{FN}$ are the
numbers of true positives, true negatives, false positives, and
false negatives. Taking both true and false positives and
negatives into account, MCC has been widely used to measure the
quality of binary classifiers. The larger the MCC is, the better
the classifier performs. Means and standard deviations of the
above measures are provided in Tables 1-3. Under each of the three
models, we generated 100 simulated data sets with different
combinations of $p$ and $n.$ We considered three scenarios:
$p=35,$ $n=100,$ and $p=75,$ $n=100.$ and $p=35,$ $n=10000.$
Specifically, when $p=35,$ and $n=100,$ the total number of
parameters in the inverse covariance matrix to be estimated is
630, which is larger than the sample size $n=100.$ When $p=75,$
and $n=100,$ the corresponding number of parameters is 2850, which
greatly exceeds $n$. Those settings can be useful to reveal the
empirical performance of the different competing methods when the
number of parameters is greater than the sample size. The settings
with $p=35,$ and $n=10000$ are used to assess the consistency of
different methods in model selection when sample size tends to
infinity.

The implementation is based on the GLASSO package in R (Friedman,
Hastie and Tibshirani, 2008) and we apply the reiterative weighted
LASSO (Fan, Feng and Wu, 2009) to to obtain the estimates for SCAD
method. For adaptive LASSO, we use the sample covariance as the
initial estimate and obtain the weights based on the sample
covariance with the power $\gamma$ set to $0.5.$

We examine the empirical performance of the three different
penalty functions under the selection of optimal tuning parameter
via BIC or cross-validation. Tables 1, 2 and 3 provide the average
number of specificity, sensitivity and Matthew's correlation
coefficient over 100 simulated data sets. Standard errors are
provided in the parenthesis. For the three cases of different
sample and matrix sizes, across all three different graphical
structures, the adaptive LASSO consistently yields better
performance than the LASSO penalty. SCAD also outperforms the
LASSO penalty except for a few cases with the sample size $n=100.$
When sample size increases, the advantages of adaptive LASSO and
SCAD are more pronounced. Very interesting to note that when
$n=10000,$ the SCAD and adaptive LASSO with BIC can yield
sensitivity and specificity close to almost 1. For AR(1), the
average specificity and sensitivity for SCAD is $0.981$ and
$1.000,$ and for adaptive LASSO are $0.965$ and $1.000.$ For
AR(2), the specificity and sensitivity for SCAD are $1.000$ and
$1.000,$ and for adaptive LASSO are $0.984,$ and $1.000.$ For
sparse graph with three edges per node, the specificity and
sensitivity for SCAD are $0.999$ and $1.000,$ and for adaptive
LASSO are $0.992,$ and $1.000.$  These results confirm with the
theoretical results that when $n$ tends to infinity, penalized
likelihood estimation with SCAD or adaptive LASSO is consistent
and selects the true graphical model with probability tending to
one. In comparison, the average specificity and sensitivity that
the penalized likelihood estimation with LASSO under BIC selection
are much lower. For instance, for AR(1) model, the sensitivity is
only about $0.721$; for AR(2) model, the sensitivity is only about
0.806. These results demonstrate that LASSO with BIC is not
consistent in model selection across different underlying
graphical structures.

We also compare the performance between BIC and 5-fold cross
validation. When sample size $n=100$, there is no complete
dominance of one tuning method over the other. For instance, under
the sparse graph with three edges per node, the relative
performance of BIC versus cross validation depends on the penalty
function. When $p=35$ and $n=100$, the overall MCC of BIC is
higher than cross validation for penalty of LASSO but lower than
cross validation for penalties of SCAD and ADAP. When $n=10000,$
BIC is more advantageous as it consistently yields higher
specificity, sensitivity, and MCC than cross validation for all
the penalties and across all the graphical models in the
simulation study. Overall, the BIC method exhibits comparable
performance as cross validation in the small sample size scenario,
but it outperforms cross validation when sample size gets large.
Computationally, BIC is more convenient to use as cross validation
is $K$ times more intensive to compute.
\par

\vskip 0.5cm

\setcounter{chapter}{4}
\setcounter{equation}{0} 
\noindent {\bf 4. Conclusion}

In this article, we investigate the tuning parameter selection for
penalized likelihood estimation of the inverse covariance matrix
in the Gaussian graphical model. We establish the consistency of
the BIC criterion to select the true graphical model with the SCAD
or adaptive LASSO penalty. Such consistency result of BIC can be
extended to the general penalized likelihood estimation problems
with these two penalties in other models satisfying mild
regularity conditions.
\par
\vskip 1cm
 \noindent {\large\bf Acknowledgment} This research is
supported by the Canadian National Science and Engineering
Research Council grant held by Gao and Wu.
\par

\newpage

\begin{landscape}
\begin{table}
 \caption{Results for AR(1) Graphical Model . Averages and standard errors from 100 runs} \label{type1}
\begin{center}
{
\begin{tabular}{lllccccccccc}
\hline\hline  p &n  & Tuning&   & $LASSO$  & & &$SCAD$  & &
&$ADAP$&
\\ \hline
\hline      &   &  & SPEC & SENS & MCC & SPEC & SENS & MCC & SPEC & SENS & MCC \\
 35 & 100   &   BIC    &   0.695   &   1.000    &   0.402    &   0.710   &   1.000    &   0.413    &   0.849   &   1.000    &   0.568  \\
    &   &       &   (0.032)   &   (0.000)    &   (0.025)    &   (0.020)   &   (0.000)    &   (0.017)    &   (0.021)   &   (0.000)    &   (0.030)  \\
    &   &   CV    &   0.620   &   1.000    &   0.348    &   0.705   &   1.000    &   0.410    &   0.824   &   1.000    &   0.533  \\
   &   &       &   (0.025)   &   (0.000)    &   (0.016)    &   (0.016)   &   (0.000)    &   (0.013)    &   (0.016)   &   (0.000)    &   (0.021)  \\
75  &100   &   BIC    &   0.791   &   1.000    &   0.362    &   0.739   &   1.000    &   0.318    &   0.867   &   0.998    &   0.453  \\
  &   &       &   (0.018)   &   (0.000)    &   (0.017)    &   (0.015)   &   (0.000)    &   (0.011)    &   (0.011)   &   (0.005)    &   (0.016)  \\
  &    &   CV    &   0.712   &   1.000    &   0.299    &   0.749   &   1.000    &   0.325    &   0.901   &   0.997    &   0.515  \\
  &    &       &   (0.017)   &   (0.000)    &   (0.012)    &   (0.006)   &   (0.000)    &   (0.005)    &   (0.007)   &   (0.005)    &   (0.015)  \\
 35 & 10000   &   BIC    &   0.721   &   1.000    &   0.424    &   0.981   &   1.000    &   0.902    &   0.965   &   1.000    &   0.839  \\
   &   &       &   (0.025)   &   (0.000)    &   (0.022)    &   (0.006)   &   (0.000)    &   (0.028)    &   (0.008)   &   (0.000)    &   (0.029)  \\
  &    &   CV    &   0.521   &   1.000    &   0.290    &   0.976   &   1.000    &   0.880    &   0.917   &   1.000    &   0.697  \\
  &    &       &   (0.030)   &   (0.000)    &   (0.016)    &   (0.007)   &   (0.000)    &   (0.031)    &   (0.017)   &   (0.000)    &   (0.040)  \\
\hline\hline
\end{tabular}%
}
\end{center}
SCAD:the SCAD penalty; LASSO: the $L_1$ penalty; ADAP:
           the adaptive LASSO penalty

\end{table}
\end{landscape}

\begin{landscape}
\begin{table}
 \caption{Results for AR(2) Graphical Model . Averages and standard errors from 100 runs} \label{type1}
\begin{center}
{
\begin{tabular}{lllccccccccc}
\hline\hline  p &n  & Tuning&   & $LASSO$  & & &$SCAD$  & &
&$ADAP$&
\\ \hline
\hline      &   &  & SPEC & SENS & MCC & SPEC & SENS & MCC & SPEC & SENS & MCC \\
 35 & 100   &   BIC    &   0.986   &   0.459    &   0.585    &   0.982   &   0.519    &   0.616    &   0.954   &   0.754    &   0.703  \\
    &   &       &   (0.013)   &   (0.113)    &   (0.055)    &   (0.016)   &   (0.114)    &   (0.050)    &   (0.029)   &   (0.135)    &   (0.051)  \\
    &   &   CV    &   0.657   &   0.960    &   0.432    &   0.812   &   0.905    &   0.554    &   0.865   &   0.910    &   0.627  \\
   &   &       &   (0.050)   &   (0.026)    &   (0.036)    &   (0.058)   &   (0.056)    &   (0.045)    &   (0.028)   &   (0.039)    &   (0.039)  \\
75  &100   &   BIC    &   0.996   &   0.382    &   0.563    &   0.995   &   0.420    &   0.579    &   0.992   &   0.486    &   0.610  \\
  &   &       &   (0.003)   &   (0.065)    &   (0.035)    &   (0.003)   &   (0.065)    &   (0.032)    &   (0.005)   &   (0.091)    &   (0.041)  \\
  &    &   CV    &   0.837   &   0.887    &   0.445    &   0.895   &   0.829    &   0.503    &   0.998   &   0.362    &   0.563  \\
  &    &       &   (0.043)   &   (0.031)    &   (0.036)    &   (0.024)   &   (0.045)    &   (0.027)    &   (0.002)   &   (0.058)    &   (0.039)  \\
 35 & 10000   &   BIC    &   0.806   &   1.000    &   0.606    &   1.000   &   1.000    &   1.000    &   0.984   &   1.000    &   0.954  \\
   &   &       &   (0.031)   &   (0.000)    &   (0.038)    &   (0.000)   &   (0.000)    &   (0.000)    &   (0.006)   &   (0.000)    &   (0.020)  \\
  &    &   CV    &   0.470   &   1.000    &   0.330    &   0.997   &   1.000    &   0.991    &   0.931   &   1.000    &   0.810  \\
  &    &       &   (0.032)   &   (0.000)    &   (0.019)    &   (0.007)   &   (0.000)    &   (0.023)    &   (0.020)   &   (0.000)    &   (0.045)  \\
\hline\hline
\end{tabular}%
}
\end{center}
SCAD:the SCAD penalty; LASSO: the $L_1$ penalty; ADAP:
           the adaptive LASSO penalty

\end{table}
\end{landscape}

\begin{landscape}
\begin{table}
 \caption{Results for a sparse Graphical Model with 3 edges per node. Averages and standard errors from 100 runs} \label{type1}
\begin{center}
{
\begin{tabular}{lllccccccccc}
\hline\hline  p &n  & Tuning&   & $LASSO$  & & &$SCAD$  & &
&$ADAP$&
\\ \hline
\hline      &   &  & SPEC & SENS & MCC & SPEC & SENS & MCC & SPEC & SENS & MCC \\
 35 & 100   &   BIC    &   0.983   &   0.460    &   0.562    &   0.992   &   0.366    &   0.538    &   0.988   &   0.458    &   0.584  \\
    &   &       &   (0.012)   &   (0.073)    &   (0.044)    &   (0.015)   &   (0.117)    &   (0.052)    &   (0.011)   &   (0.092)    &   (0.053)  \\
    &   &   CV    &   0.988   &   0.363    &   0.546    &   0.998   &   0.354    &   0.558    &   0.995   &   0.423    &   0.591  \\
   &   &       &   (0.067)   &   (0.095)    &   (0.049)    &   (0.003)   &   (0.059)    &   (0.042)    &   (0.003)   &   (0.060)    &   (0.051)  \\
75  &100   &   BIC    &   0.992   &   0.416    &   0.539    &   0.997   &   0.339    &   0.534    &   0.995   &   0.389    &   0.541  \\
  &   &       &   (0.003)   &   (0.040)    &   (0.032)    &   (0.006)   &   (0.072)    &   (0.026)    &   (0.003)   &   (0.045)    &   (0.030)  \\
  &    &   CV    &   0.998   &   0.353    &   0.555    &   0.998   &   0.353    &   0.549    &   1.000   &   0.303    &   0.536  \\
  &    &       &   (0.001)   &   (0.030)    &   (0.027)    &   (0.001)   &   (0.030)    &   (0.023)    &   (0.000)   &   (0.024)    &   (0.019)  \\
 35 & 10000   &   BIC    &   0.932   &   1.000    &   0.769    &   0.999   &   1.000    &   0.996    &   0.992   &   1.000    &   0.966  \\
   &   &       &   (0.017)   &   (0.000)    &   (0.044)    &   (0.002)   &   (0.000)    &   (0.009)    &   (0.004)   &   (0.000)    &   (0.019)  \\
  &    &   CV    &   0.657   &   1.000    &   0.407    &   0.997   &   1.000    &   0.991    &   0.959   &   1.000    &   0.851  \\
  &    &       &   (0.045)   &   (0.000)    &   (0.034)    &   (0.003)   &   (0.000)    &   (0.017)    &   (0.029)   &   (0.000)    &   (0.081)  \\
\hline\hline
\end{tabular}%
}
\end{center}
SCAD:the SCAD penalty; LASSO: the $L_1$ penalty; ADAP:
           the adaptive LASSO penalty
\end{table}
\end{landscape}

\noindent{\large\bf References}
\begin{description}

\item Banerjee, O., Ghaoui, L. E. and D'Aspremont, A. (2007).
Model selection through sparse maximum likelihood estimation.
\emph{Journal of Machine Learning Research} \textbf{9}, 485-516.

\item Dempster, A. P. (1972). Covariance selection.
\emph{Biometrika} \textbf{32}, 95-108.

\item Edwards, D. M. (2000). \emph{Introduction to Graphical
Modelling.} New York: Springer.

\item Fan, J. and Li, R. (2001). Variable selection via nonconcave
penalized likelihood and its oracle properties. \emph{Journal of
the American  Statistical Association} \textbf{96}, 1348-60.

\item Fan, J., Feng, Y. and Wu, Y. (2009). Network exploration via
the adaptive LASSO and SCAD penalties. \emph{The Annals of Applied
Statistics} \textbf{3}, 521-541.

\item Friedman, J., Hastie, T. and  Tibshirani, R. (2008). Sparse
inverse covariance estimation with the graphical lasso.
\emph{Biostatistics} \textbf{9}, 432-441.

\item Li, H. and  Gui, J. (2008). Gradient directed regularization
for sparse gaussian concentration graphs, with applications to
inference of genetic networks. \emph{Biostatistics} \textbf{7},
302-317.

\item Lauritzen, S. L. (1996). \emph{Graphical Models}. Oxford:
Clarendon Press.

\item Lehmann, E. L. (1983). \emph{Theory of Point Estimation}.
Pacific Grove, CA: Wadsworth and Brooks/Cole.

\item Meinshausen, N. and B\"{u}hlmann, P. (2006).
High-dimensional graphs with the Lasso. \emph{Ann. Statist.}
\textbf{34}, 1436-62.

\item Tibshirani, R. J. (1996) Regression shrinkage and selection
via the lasso. \emph{Journal of the Royal Statistical Society,
Series B}, \textbf{58}, 267-288.

\item Wang, H., Li, R. \& Tsai, C.-L. (2007). Tuning parameter
selectors for the smoothly clipped absolute deviation method.
\emph{Biometrika}. \textbf{94}, 553-68.

\item White, H. (1982) Maximum Likelihood Estimation of
Misspecified Models. \emph{Econometrika}, 50, 1-25.

\item Whittaker, J. (1990) \emph{Graphical Models in Applied
Multivariate Statistics}. Chichester: John Wiley and Sons.

\item Yuan, M. \& Lin, Y. (2007). Model selection and estimation
in the Gaussian graphical model. \emph{Biometrika} \textbf{94},
19-35.

\item Zou, H. (2006) The adaptive Lasso and its oracle properties.
\emph{Journal of the American  Statistical Association}
\textbf{101}, 1418-1429.

\item Zou, H. and Li, R. (2008) One-step sparse estimates in
nonconcave penalized likelihood models (with discussion).
\emph{Annals of Statistics} , \textbf{36}, 1509-1533.

\end{description}

\vskip .65cm \noindent Xin Gao, Department of Mathematics and Statistics\\
York University, Toronto, ON\vskip 2pt \noindent E-mail:
xingao@mathstat.yorku.ca \\
Tel: 416-736-2100 ext66097\\
Fax:416-736-5757 \vskip 2pt \noindent Daniel, Q. Pu, Department of
Mathematics and Statistics\\ York University, Toronto, ON \vskip
2pt \noindent E-mail: puq@mathstat.yorku.ca \vskip 2pt \noindent
Yuehua Wu, Department of Mathematics and Statistics\\ York
University, Toronto, ON \vskip 2pt \noindent E-mail:
wuyh@mathstat.yorku.ca
\vskip 2pt \noindent Hong Xu, Department of Mathematics and Statistics\\
York University, Toronto, ON \vskip 2pt \noindent E-mail:
hongxu@mathstat.yorku.ca \vskip .3cm
\end{document}